\documentclass{article}

\usepackage{arxiv}

\usepackage[utf8]{inputenc} 
\usepackage[T1]{fontenc}    
\usepackage{hyperref}       
\usepackage{url}            
\usepackage{booktabs}       
\usepackage{amsfonts}       
\usepackage{nicefrac}       
\usepackage{microtype}      
\usepackage{lipsum}
\usepackage{graphicx}
\usepackage{color}
\usepackage{tablefootnote}
\usepackage{multirow}
\graphicspath{ {./images/} }

\title{Neural News Recommendation with Event Extraction}

\author{
 Songqiao Han\textsuperscript{$\ast$} 
   \And
 Hailiang Huang\textsuperscript{$\ast$} 
  \And
 Jiangwei Liu\thanks{All authors contribute equally.} 
 \And
 \\
 School of Information Management and Engineering,\\
 Shanghai University of Finance and Economics, Shanghai 200433, China \\
}

\begin{document}
\maketitle

\begin{abstract}
A key challenge of online news recommendation is to help users find articles they are interested in. Traditional news recommendation methods usually use single news information, which is insufficient to encode news and user representation. Recent research uses multiple channel news information, e.g., title, category, and body, to enhance news and user representation. However, these methods only use various attention mechanisms to fuse multi-view embeddings without considering deep digging higher-level information contained in the context. These methods encode news content on the word level and jointly train the attention parameters in the recommendation network, leading to more corpora being required to train the model. We propose an Event Extraction-based News Recommendation (EENR) framework to overcome these shortcomings, utilizing event extraction to abstract higher-level information. EENR also uses a two-stage strategy to reduce parameters in subsequent parts of the recommendation network. We train the Event Extraction module by external corpora in the first stage and apply the trained model to the news recommendation dataset to predict event-level information, including event types, roles, and arguments, in the second stage. Then we fuse multiple channel information, including event information, news title, and category, to encode news and users. Extensive experiments on a real-world dataset show that our EENR method can effectively improve the performance of news recommendations. Finally, we also explore the reasonability of utilizing higher abstract level information to substitute news body content.
\end{abstract}


\section{Introduction}
With the rapid development of social media, online news platforms receive and generate massive news every day, which leads to potential information overloading problems. Recommendation systems play a crucial role in helping users target the expected news among the mess. There have been various approaches proposed for personalized news recommendation, which could be divided into three typical categories: collaborative filtering (CF) based methods \cite{XueDai-44,WangHe-47}, content-based methods \cite{WangZhang-27,WuWu-42,HuLi-23} and Hybrid methods. However, CF-based methods suffer from data sparsity problems because news is out-of-date and substituted frequently \cite{HuLi-23}. And recent content-based methods focus on multiple views to exploit the data but ignore exploring deep into the text content. For example, literature \cite{WuWu-42,HuLi-23} utilized attention mechanisms on words to select important information, which may be insufficient to utilize higher-level information contained in the body content.

When we try to answer "What Do News Readers Really Want to Read About?", we find that people favor news events that may affect them and their lives. Our motivation is to extract higher-level information than word-level, e.g., news events, to encode news and user representation better. With the exponential growth of digital natural language text, e.g., news messages, research in event extraction (EE), one of the core Information Extraction (IE) tasks, has significantly evolved. For example, \cite{BorsjeHogenboom-57} leverages lexico-syntactic patterns to a higher abstraction level. Then, these new lexico-syntactic patterns are used to generate rules on financial ontologies to discover financial events. Reference \cite{BalaliAsadpour-58} proposed an attention-based graph convolutional network to extract event triggers and arguments simultaneously by introducing Shortest Dependency Path (SDP) in the dependency graph. Compared with the domain-rules based EE approaches, deep learning-based EE approaches have better versatility. The disadvantage of deep learning based EE approaches is that they rely on huge annotated data, which involves a time-consuming and laborious process. We balance the advantages and disadvantages, adopt a similar external domain annotated corpus to alleviate the gap. Due to Event Extraction could answer who did what, when, where, and why, it abstracts the core idea of contents. The extracted results may be regarded as an alternative representation of the content, avoiding introducing more noise than that all words fully participate.

In this paper, we propose an Event Extraction based News Recommendation (EENR) framework. The main idea is to utilize event extraction technology to enhance recommendation performance. Unlike previous research, we use event extraction technology to dig deep into news content and obtain event types, roles, and arguments. Event-level information helps better encode news and user representation. Especially, we utilize attention mechanisms to combine homogeneous information, for example, integrating semantic information to represent news and users. While dealing with multi-channel heterogeneous information, we concatenate event type embedding, semantic embedding, and category embedding sequentially to represent news and users, respectively. Experimental results on real-world news recommendation datasets demonstrate that our model outperforms the most used and state-of-the-art methods.

The contributions of this paper are summarized as follow.
\begin{itemize}
	\item  We propose using event extraction to abstract higher-level information contained in news content to enhance news and user representation. It avoids introducing more noise than that all words participate. And our event extraction module is individually trained on similar public domain annotated data. It avoids introducing more attention parameters in the subsequent modules of the recommendation system and ensures it works well on small recommendation data sets.
	
	\item  An EENR framework is proposed for news recommendation, fusing multi-channel information (e.g., event types, roles, arguments, news titles, news categories). EENR utilizes external knowledge and event extraction technologies to provide additional views to encode news and user representation, which enriches the general recommendation frameworks. Lastly, we also explore the reasonability of utilizing higher abstract level information to substitute news body content. 
\end{itemize}

The remainder of this paper is organized as follows. We review related work in Section 2 and propose our EENR framework in Section 3. Section 4 reports experimental results and discusses the reasonability, followed by the conclusion in Section 5.

\section{Related Work}
\subsection{News Recommendation}
Personalized news recommendation has become a pervasive feature of online news platforms and has been widely studied. Various methods have been proposed and could be divided into three typical categories: collaborative filtering (CF) based methods, content-based methods, and hybrid methods.

CF methods predict a personalized ranking over a set of items for each user with the similarities among the users and items \cite{XueDai-44,WangHe-47}. For example, \cite{WangHe-47} proposed a neural graph collaborative filtering (NGCF) framework to exploit the high-order user-item interaction information. However, these methods still suffer sparsity of user-item interactions for news is substituted frequently with time.

Content-based methods \cite{WangZhang-27, WuWu-42, ZhuZhou-29} mainly concern the semantic information of news. Much previous work concerned a single kind of information, but recently, researchers exploited different kinds of news information. For example, \cite{WangZhang-27} fused semantic-level and knowledge-level representation of news through DKN - a multi-channel and word-entity-aligned convolutional neural network. Reference \cite{WuWu-42} used attention mechanisms to fuse multi-channel information: title, category, and body. Reference \cite{WangZhang-50} proposed a knowledge-aware graph neural network with label smoothness regularization to capture semantic relationships between items. However, these methods focused on neural network designing and still ignored higher abstract level information contained in news content, e.g., event-level information.

Hybrid methods \cite{LenhartHerzog-49,YeTu-48} usually integrate different recommendation algorithms.

Our EENR model follows content-based methods. In recent research, news bodies are utilized roughly by weighting the word embeddings, which may consume more parameters and introduce semantic noises. We inspect the process of how neurons work before people click certain news, argue that the "content" before people click the news is abstract and higher-level, not word-level in detail. Thus event-level information is more suitable as model input. EENR aims to solve this problem and thus introduces an event extraction module. The experiment results demonstrate the effectiveness of our idea.

\subsection{Event Extraction}
Event Extraction is an advanced form of Information Extraction that handles textual content or relations between entities, often performed after executing a series of initial NLP steps \cite{HogenboomFrasincar-53}. Event extraction from news articles is a commonly required prerequisite for various tasks, such as article summarization, article clustering, or news aggregation \cite{HamborgBreitinger-51}. From the view of technical implementation maps, event extraction approaches include domain-rules based methods, deep learning based methods, and hybrid methods. For instance, \cite{HamborgBreitinger-51} improved the universal EE model (Giveme5W1H) to detect main events in an article. It used syntactic and domain-specific rules to extract the relevant phrases to answer 5W1H questions automatically: who did what, when, where, why, and how. Reference \cite{XuLiu-52} proposed a hybrid collaborative filtering method for event recommendation in event-based social networks. Reference \cite{LuVoss-55} developed a comprehensive system that searched, identified, and summarized complex events from multiple data modalities, then recommended events related to the user's ongoing search. However, these previous works mainly focused on recommending related events or information similar to the same event. There is limited research on utilizing event extraction to facilitate news recommendation.

Different from the above work, our approach contains a deep neural Event Extraction module, which conducts event extraction on news bodies and outputs news arguments, roles, and event types, to enhance news and user representation. In detail, event arguments can be regarded as keywords reflecting the main idea; event roles could also be regarded as external knowledge tags enhancing the intelligibility of arguments, and event types provide another perspective to inspect the news. From another view, to inspect the role of the event extraction module in the EENR framework, event roles and arguments can be regarded as an approximative substitution of the news body. We discuss the reasonability in the subsequent section in detail.

\section{Our Approach}
\begin{figure}[htbp]
	\centering
	\includegraphics[width= 6.5 in]{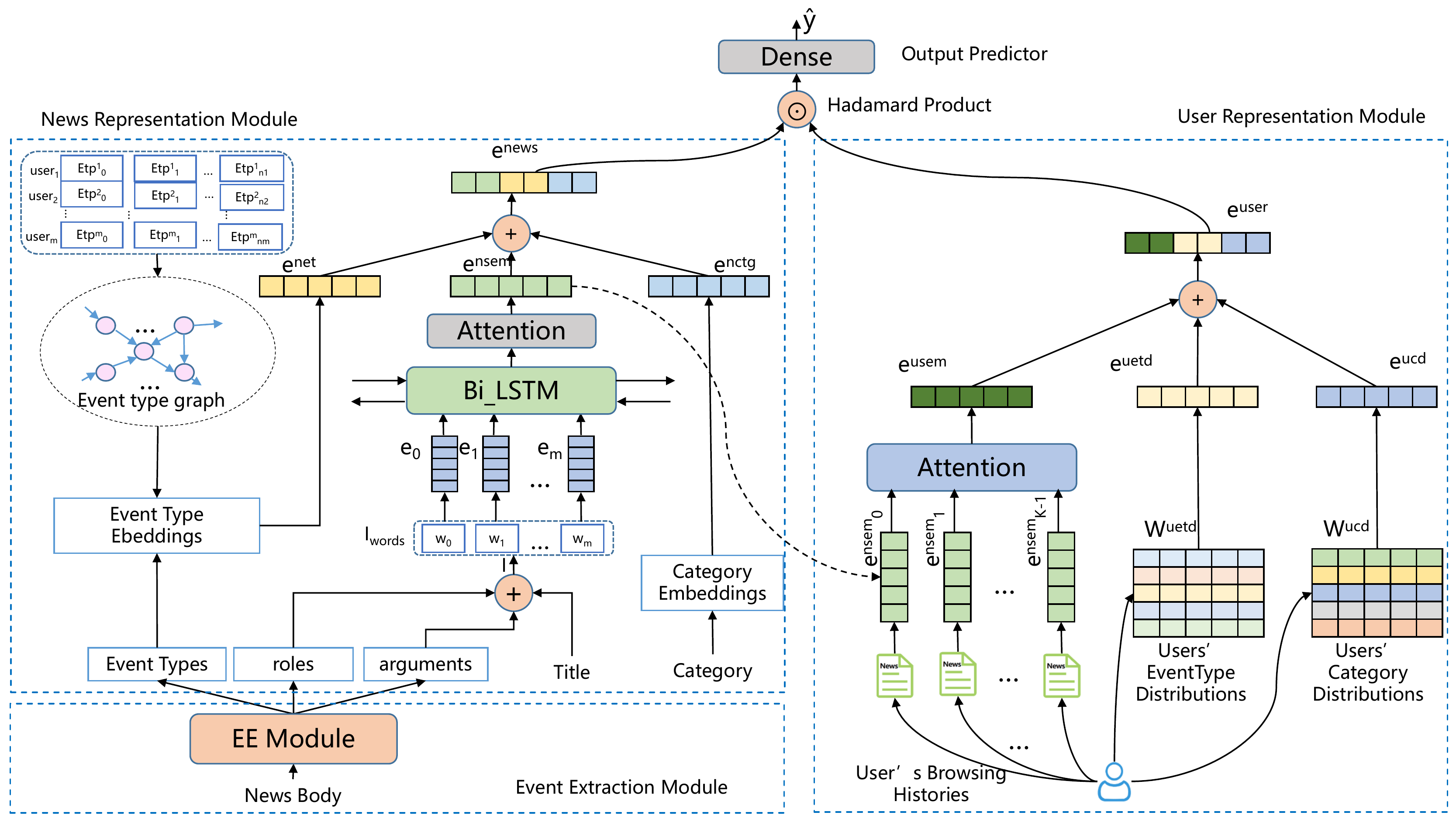}	
	\caption{The framework of EENR.}
	\label{fig1}	
\end{figure}

In this section, we introduce our neural news recommendation framework EENR. As illustrated in Figure \ref{fig1}, there are four main components: event extractor module extracts news arguments, roles, and event types from news body; news representation module encodes news from multi-channel information; user representation module encodes users from their browsing histories; output predictor module combines all input embeddings to predict a probability for one user to click candidate news.

\subsection{Event Extractor}
Traditional event extraction approaches have some drawbacks: domain-rules based methods relying on elaborately designed features while deep learning based methods relying on huge manually annotated data. Due to most EE tasks are domain-specific, so existing shared elaborately designed features and manually annotated data cannot be reused directly. Some recent research focuses on automatic EE methods, e.g., \cite{HuangJi-56}. However, these methods still have limitations. For instance, extracted event roles or types may deviate far from expected. It may get significantly worse in the news recommendation scenario because massive new news keeps emerging. Considering that deep learning based EE methods have a better generalization and domain-rules based methods are well targeted in specific fields, we designed an EE module taking advantage of both to extract event information from news bodies. Two strategies are adopted to overcome the challenges mentioned above.

First, considering similar domain event extraction corpora have similar event type distributions, we utilize a similar shared domain annotated dataset to train the EE module. It helps overcome two problems: lack of annotated data and extracted event roles or types deviating. Our target news recommendation dataset is masking log files recorded from one popular Chinese securities APP. After manually checking the content, it contains various categories except for financial news, such as "DISASTER/ACCIDENT" news. We choose the Chinese Event Extraction Dataset adopted in Language and Intelligent Technology Competition 2020 (DuEE 2020) to train the EE module for two reasons. On the one hand, DuEE corpus is selected from the hot search board of Baidu, which reflects various interests of most Chinese people; thus, it is consistent with our dataset distributions to a large extent. On the other hand, this well-annotated dataset would alleviate the roles and event types drifting problem.

\begin{figure}[htbp]
	\centering
	\includegraphics[width= 4.5 in]{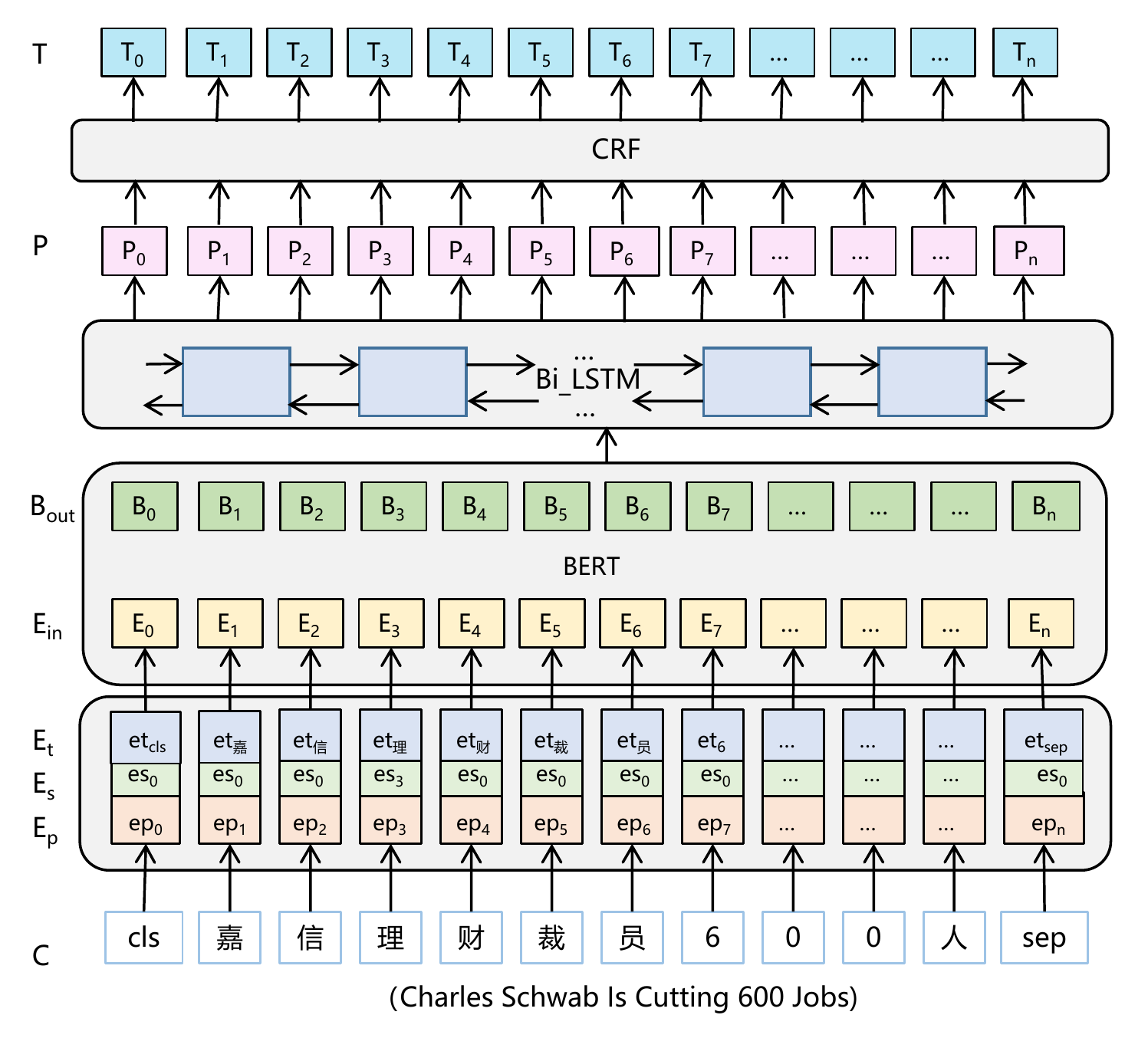}	
	\caption{The architecture of Event Extractor module.}
	\label{fig2}	
\end{figure}

Second, considering the generalization of deep learning based models, we design and train a deep neural network and apply it to our target news recommendation dataset. We also utilize some simple rules to filter and modify the neural network output. Motivated by the idea of text classification, we transform the recognition of event types and event elements into a classification task. The architecture of the EE module is illustrated in Figure \ref{fig2}.

The character-based inputs are denoted as $C = [{c_0},{c_1},...,{c_n}]$ , where $n$ is character length of the sentence. Bert module considers the corresponding token embeddings ${E_t} = [e_{c0}^t,e_{c1}^t,...,e_{cn}^t]$, segment embeddings ${E_s} = [e_0^s,e_0^s,...,e_0^s]$, and position embeddings ${E_p} = [e_{c0}^p,e_{c1}^p,...,e_{cn}^p]$. We combine these features as Bert module's internal input, denoted as ${E_{in}} = [{E_0},{E_1},...,{E_n}]$. Bert module learns the parameters mainly through transfer learning and outputs corresponding character embeddings ${B_{out}} = [{B_0},{B_1},...,{B_{\rm{n}}}]$. Then BiLSTM layer is used to learn sequence information, considering both the past and the future context of the tokens. The Softmax function is applied to the output vector of BiLSTM to normalize the probabilities on each sub-tag. For example, suppose ${P_{\rm{i}}}$ is a probability vector $[0.6,0.1,0.15,0.0,...,0]$, $0.6$ denotes this character has a probability of $60\%$ at the beginning of \emph{"layoff executor"} phrase, $0.1$ denotes that it has a probability of $10\%$ in the middle of \emph{"layoff executor"} phrase, and so on. The CRF layer introduces some constraints (e.g., a transition score matrix for each transition pattern between adjacent tags) to model correlations between adjacent tags, thus leading to a better result. The final structured output list of the example in Figure \ref{fig2} looks like: 
\begin{itemize}
	\item  \emph{[{"event type": "Organizational-Relations/Layoff", "arguments": [{"role": "number of job cut", "argument": "600"}, {"role": "layoff executor", "argument": "Charles Schwab"}], "class": "Organizational-Relations"}].}
\end{itemize}

The EE module is trained on DuEE dataset and then applied to the target news recommendation dataset.

\subsection{News Representation}
The News Representation module learns news representation from multi-channel information, such as title, category, and the output of the EE module. As illustrated in the upper left-hand part of Figure \ref{fig1}, it has five steps to process the information.

The first step is a prerequisite for the semantic representation, fusing event role, argument, and news title. The first intuition is adopting attention mechanisms to handle the parallel channel information and feed it to BiLSTM. But our experiment results in this step show that concatenating title, event arguments, and roles in sequence is a preferable choice. Thus final input $I$ for BiLSTM is denoted as:
\begin{equation} 
	I{\rm{ }} = {\rm{ }}\left[ {title;{\rm{ }}argument;{\rm{ }}role} \right] \label{eq1}
\end{equation}

The second step is responsible for the semantic representation of news. The Chinese segmentation word list eliminating unknown words is denoted as ${I_{words}} = [{w_0},{w_1},...,{w_m}]$, its corresponding word embedding is represented as $e = [{e_0},{e_2},...,{e_m}]$. BiLSTM incorporates a forward LSTM layer and a backward LSTM layer to learn information from both preceding and following tokens. We combine forward final hidden states and backward final hidden states by weighting parameters. The final semantic information of news is formulated as:
\begin{equation} 
	{e^{nsem}} = \alpha *fwd\_states + (1 - \alpha )*bwd\_states \label{eq2}
\end{equation}

The third step is constructing an event type graph to learn event type embeddings. Given a user $use{r_m}$, the historical event type sequence is denoted as $[Etp_0^m,Etp_1^m,...,Etp_{nm}^m]$. Event types are selected as nodes in the event type graph shown in the upper left-hand corner in Figure \ref{fig1}. We use Node2Vector to train this graph to get event-type embeddings. They are initial values for the subsequent steps and would be fine-tuned during the recommendation training process.

The fourth step initiates category embeddings. Different from the third step, category embeddings are initiated by semantic meaning. Considering that news categories are well classified by humans, we use semantic embeddings as their initial values. Same as the third step, category embeddings can also be fine-tuned later.

The last step combines the previous channel information. Those categorical and semantic embeddings are obtained separately through different techniques, measured, and recorded in different spatial coordinates. So attention mechanism is not very suitable for combining these embeddings straightly. Finally, to avoid straight intersection of heterogeneous information, news representation is organized by concatenating news semantic embeddings, event type embeddings, and category embeddings in sequence, which is formulated as:
\begin{equation} 
	{e^{news}} = [{e^{nsem}},{e^{net}},{e^{nctg}}] 
	\label{eq3}
\end{equation}

Middle experiment results also demonstrate that it is a preferable choice than straightly using an attention mechanism to combine the heterogeneous information.

\subsection{User Representation}
Similar to the News Representation module, user representation needs three steps to generate user embeddings.

The first step is generating users' semantic representations. Existing literature has shown that the latest browsing records are more valuable to predict a user's subsequent behaviors. However, some users have several times as many records as the others, in which case the predictor tends to favor the users having more records. To avoid this unbalanced case, referring to the methodology \cite{ZhuZhou-29}, we select a fixed length of recent browsing histories for each user representation. Given that a user has $K$ recent records $[new{s_0},new{s_1},...,new{s_{K - 1}}]$, their semantic representation $[e_0^{nsem},e_1^{nsem},...,e_{K - 1}^{nsem}]$ can be generated by the news representation module. We use attention mechanisms to select the import records. The importance score of the $ith$ news is denoted as ${a_i}$, with its normalized form ${\alpha _i}$. The attention process of selecting important news is formulated as:
\begin{equation} 
\begin{array}{l}
	{a_i} = {q^T}tanh(W \times e_i^{nsem} + b)\\
	{\alpha _i} = \frac{{exp({a_i})}}{{\sum\nolimits_{j = 0}^{K - 1} e xp({a_i})}}
\end{array} 
\label{eq4}
\end{equation}

where transform matrix $W$, query vector $q$, and bias vector $b$ are attention parameters to learn. The user's semantic representation is a weighted combination of his historical news, which is computed by:
\begin{equation} 
	{e^{usem}} = \sum\nolimits_0^{K - 1} {{\alpha _i}} e_i^{nsem}
	\label{eq5}
\end{equation}

The second step generates the user's event type distribution by looking up in event type distribution matrix ${W^{uetd}}$. ${W^{uetd}}$ are parameters to learn and initiated by averaging users' event type embeddings. One's event type distribution is acquired by querying ${W^{uetd}}$ by user id, denoted as ${e^{uetd}}$.

Same as the above, one's category distribution embedding ${e^{ucd}}$ can be looked up in users' category distribution matrix ${W^{ucd}}$ by user id. In agreement with news representation, the user representation is also organized by concatenating one's semantic embedding, event type distribution, and category distribution in sequence, which is formulated as:
\begin{equation} 
	{e^{user}} = [{e^{usem}},{e^{uetd}},{e^{ucd}}]
	\label{eq6}
\end{equation}

\subsection{Output Predictor}
Once the news and users are encoded to embeddings, we utilize user representation and news representation to predict a probability score $\hat y$ that users click the candidate news. Unlike previous research \cite{GeWu-24} that predicted clicking probability by using the Inner Product of news and user representation, we use a DNN network whose input is element-wise (Hadamard) product of news and user representation. First, the news representation ${e^{news}}$ and user representation ${e^{user}}$ are combined as:
\begin{equation} 
	{e^{nu}} = {e^{news}} \odot {e^{user}}
	\label{eq7}
\end{equation}

Where $\odot$ denotes the element-wise (Hadamard) product. A DNN layer and normalized layer are followed to calculate the probability score $\hat y$. Finally, normalized probabilities grouped by the user are ranked, and then the output predictor module returns the user's suggestion list.

\section{Experiments}
\subsection{Datasets and Experimental Settings}
In this section, we first introduce two datasets: Chinese Event Extraction Dataset and News Recommendation Dataset. Chinese Event Extraction Dataset is used to train EE modular, and News Recommendation Dataset is our target recommendation dataset. The experimental settings and evaluation matrics are followed.

\subsubsection{Chinese Event Extraction Dataset}
Baidu released the Chinese Event Extraction Dataset \footnote[1]{http://lic2020.cipsc.org.cn/} (DuEE), adopted in Language and Intelligent Technology Competition 2020. The corpus was selected and determined according to the hot search board of Baidu. It consists of 17,000 sentences containing 20,000 events of 65 event types. The event types cover the traditional event extraction types, e.g., “earthquake”, and various financial and economic event types, e.g., “limit down”. The copus format is as follow. 
\begin{itemize}
	\item  \emph{{“id”: “id1”, “event list”: [{“event type”: “T1”, “arguments”: [{“role”: “R1”, “argument”: “A1”}, …]}, {“event type”: “T2”, “arguments”: [{“role”: “R2”, “argument”: “A2”}, …]}]}.}
\end{itemize}
 The event Extraction module was trained on this dataset and then was applied to extract event-level information to the target news recommendation dataset. After being checked and merged, there are about 27 event classes left on the recommendation dataset.

\subsubsection{News Recommendation Dataset} 
Our target news recommendation dataset is masking log files of one Chinese securities APP collected from Oct. 30th, 2019 to Nov. 7th, 2019. The news dataset contains 10976 users, 14909 news, and 554796 sessions. We adopt the 7:1:2 proportion to split it into train, validation, and test set by time.

\subsubsection{Experimental Settings}
We implement our model based on Tensorflow. The performance on the validation set determined the hyperparameters. Negative sampling ratio K was set to 4 empirically according to \cite{WuWu-42}. The dimension of pre-trained word embedding is 300. Categorical embeddings, including original news category embedding, extracted event type embedding, user's category distribution embedding, and user's event type distribution embedding, were all set to 50. The batch size was set to 128. Other hyper-parameter settings follow previous works \cite{WuWu-42,HuLi-23}. We use impression-based ranking metrics to evaluate the performance in agreement with previous research, including area under the ROC curve (AUC), mean reciprocal rank (MRR), and normalized discounted cumulative gain (NDCG).

\subsection{Baselines and Performance Evaluation}
We choose the following most used and state-of-the-art methods as baselines in our experiments: 

\begin{itemize}
	\item \textbf{LibFM} \footnote{http://www.libfm.org/} \cite{rendle-tist2012} is a most used implementation for factorization machine that features Bayesian inference using Markov Chain Monte Carlo (MCMC).
	
	\item \textbf{KimCNN} \cite{Kim-45} achieves excellent results on multiple benchmarks, including sentiment analysis and text classification.
	
	\item \textbf{Wide\&Deep} \cite{ChengKoc-46} is a jointly trained wide linear and deep nonlinear neural network. This architecture integrates the benefits of memorization (e.g., wide linear models) and generalization (e.g., deep neural networks).
	
	\item \textbf{DFM} \cite{XueDai-44} is a matrix factorization model with neural network architecture. DFM learns a common low dimensional space to represent users and items without considering news contents or other auxiliary information.
	
	\item \textbf{NAML} \cite{WuWu-42} encodes news and user representation by different kinds of news information, such as title, body, and category. Attention mechanisms integrate the multi-view information.
\end{itemize}

Our method EENR and its four variants are listed as follows: 
\begin{itemize}
	\item \textbf{TITLE}: we only consider the title information.
	
	\item \textbf{TITLE+RA}: we consider the title concatenated with roles and arguments of extracted events.
	
	\item \textbf{TITLE+ET+RA}: we consider the title, event type, roles, and arguments of extracted events. The difference between TITLE+ET+RA and EENR lies in whether considering the original news category.
	
	\item \textbf{TITLE+NT+RA}: it is the same as TITLE+ET+RA except that TITLE+NT+RA utilizes original category embedding while TITLE+ET+RA uses event type embedding.
	
	\item \textbf{EENR}: our proposed method uses event extraction to enhance the performance of news recommendation; it considers title, original news category, event type, roles, and arguments of extracted events.
\end{itemize}
	
For the news dataset having no ground-truth labels of event extraction, we randomly sample 300 news articles to evaluate the performance manually. The statistical macro-average precision and AUC are 0.76 and 0.80, respectively. This favorable performance promises that the EE component is qualified for providing reliable event information for the subsequent modules of EENR.

We do our best to optimize the parameters of the listed models, then conduct experiments to compare our model with the benchmarks. We repeat every model 10 times and report the average results in Table \ref{tab1}.

\begin{table}
	\caption{The performance scores of different methods.}
	\centering
	\begin{tabular}{lcccc}
		\toprule
		Methods & AUC	& MRR	& NDCG@5	& NDCG@10\\
		\midrule
		LibFM	& 0.7227	& 0.4672	& 0.4590	& 0.4633\\
		KimCNN	& 0.8129	& 0.5507	& 0.5476	& 0.5714\\
		Wide\&Deep	& 0.8566	& 0.6081	& 0.6216	& 0.6431\\
		DFM	& 0.7797	& 0.4567	& 0.4560	& 0.4791\\
		NAML	& 0.8617	& 0.6128	& 0.6246	& 0.6460\\
		\midrule
		TITLE	& 0.8458	& 0.5345	& 0.5490	& 0.5772\\
		TITLE+RA	& 0.8437	& 0.5419	& 0.5553	& 0.5825\\
		TITLE+ET+RA	& 0.8688	& 0.6348	& 0.6504	& 0.6681\\
		TITLE+NT+RA	& 0.8731	& 0.6344	& 0.6487	& 0.6672\\
		EENR	& 0.8764	& 0.6491	& 0.6629	& 0.6808\\
		\bottomrule
	\end{tabular}
	\label{tab1}
\end{table}

From Table \ref{tab1}, we have several observations:

First, content-based methods (e.g. CNN, Wide\&Deep, NAML, EENR) outperform matrix factorization based models (e.g. DFM, LibFM). The reason probably is that content-based methods utilize semantic embedding, such as word embedding, which is trained on huge corpora and thus contains external knowledge. The external knowledge may help models learn better news and user representation, while factorization models learn implicit representation, which is not straightforward and thus sub-optimal.

Second, the methods considering more channel information of news (e.g., EENR, NAML) outperform those with less channel information (e.g., TITLE, TITLE+RA). The reason is that each channel information is complementary and could afford different views to the model; thus, integrating them could lead to better performance. Especially, TITLE+ET+RA outperforms TITLE by a significant margin, showing the effectiveness of Event Extraction module. 

Third, EENR achieves the best result. Compared with the state-of-the-art NAML method, EENR surpasses 1.1\%, 3.6\%, 3.8\%, 3.4\% in AUC, MRR, NDCG@5, and NDCG@10. The excellent performance mainly can be attributed to Event Extraction module due to its two advantages: event type introduces external knowledge implicitly; abstract higher-level information avoids introducing more noise.

\subsection{Effectiveness of Event Extraction}
Except for the two advantages we analyzed above, another advantage of Event Extraction module is that it has superiority on smaller datasets. In this section, we conduct more experiments on these three scenarios to verify the effectiveness of event extraction component.

\paragraph{(1)}  Event extraction module outputs event types, which offer another perspective to study a person's interest distribution. For example, from the original news category view, one's top interest includes bulletin, stock comments, expert comments, and securities research reports. In contrast, from the event type view, his interest could be described as deal-sell/buy, deal-limit down, organizational relationship-join, product behavior-release. This complementary information may help fully characterize one's interest points and thus enhance the recommendation performance. The performance of models TITLE+ET+RA and TITLE+RA shows that event type averagely brings 9.3\% lift in MRR and 8.6\% lift in NDCG@10, which demonstrates event type plays an important complementary role.

\begin{figure}[htbp]
	\centering
	\includegraphics[width= 5in]{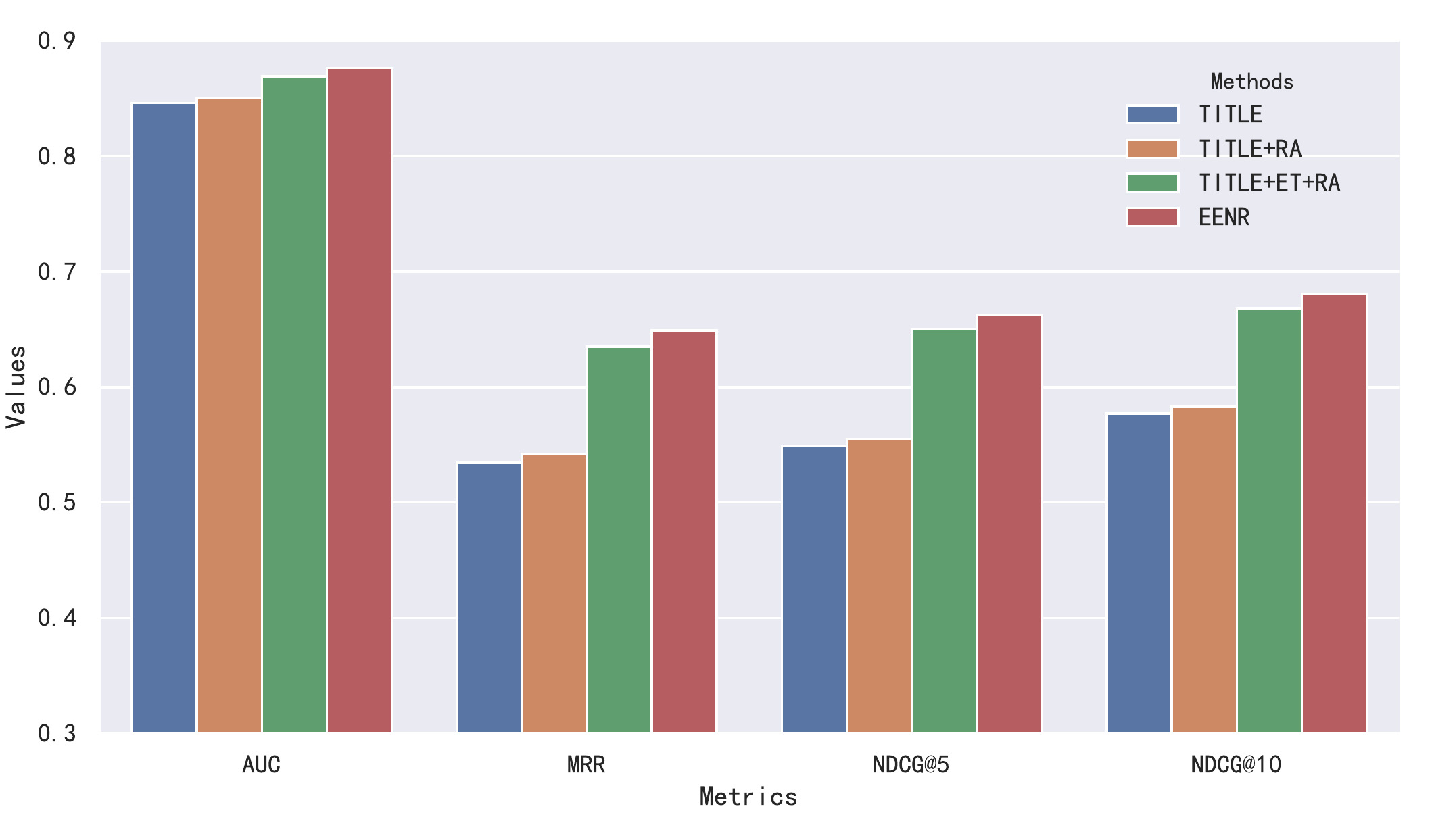}	
	\caption{Illustration of event effectiveness.}
	\label{fig3}	
\end{figure}

\paragraph{(2)}  We argue that event roles and arguments contain the main ideas and can substitute news bodies. For a fair comparison, we fixed other parts and designed a model TITLE+ET+RA based on NAML. In terms of AUC, MRR, NDCG@5, and NDCG@10, TITLE+ET+RA obtains substantially better average improvement than NAML 1.2\%, 2.2\%, 2.4\%, and 2.1\%, respectively. Although attention mechanisms would select the more important words in the body, total words participate is inferior to only event roles and arguments. The experimental results provide support for our proposition that higher abstract level features could avoid introducing more noises. More ablation experiments give evidence to support the two points above, and the results are plotted in Figure \ref{fig3}. For example, the MRR and NDCG@10 of the TITLE+ET+RA model surpass the TITLE model by 10.0\% and 9.1\%, respectively.

\begin{table}
	\caption{The performance scores of different methods on smaller datasets. P=0.2 and p=0.6 denote 20\%, 60\% of the total dataset are used, respectively.}
	\centering
	\begin{tabular}{lllllllll}
		\toprule
		\multirow{2}{1 in}{Methods} & \multicolumn{2}{c}{AUC} & \multicolumn{2}{c}{MRR} & \multicolumn{2}{c}{NDCG@5} & \multicolumn{2}{c}{NDCG@10}\\
		& p=0.2	& p=0.6	& p=0.2	& p=0.6	& p=0.2	& p=0.6	& p=0.2	& p=0.6\\
		\midrule
		TITLE	& 0.7927 	& 0.8092 	& 0.4535 	& 0.4773 	& 0.4680 	& 0.4901 	& 0.4974 	& 0.5238 \\
		TITLE+RA	& 0.8055 	& 0.8123 	& 0.4672 	& 0.4883 	& 0.4733 	& 0.5017 	& 0.4975 	& 0.5290 \\
		TITLE+ET+RA	& 0.8197 	& 0.8324 	& 0.5321 	& 0.5739 	& 0.5475 	& 0.5885 	& 0.5680 	& 0.6096 \\
		NAML	& 0.8118 	& 0.8402 	& 0.4956 	& 0.5526 	& 0.5040 	& 0.5638 	& 0.5327 	& 0.5900 \\
		EENR	& 0.8337 	& 0.8581 	& 0.5676 	& 0.6054 	& 0.5829 	& 0.6146 	& 0.6016 	& 0.6380 \\		
		\bottomrule
	\end{tabular}
	\label{tab3}
\end{table}

\begin{figure}[htbp]
	\centering
	\includegraphics[width= 6 in]{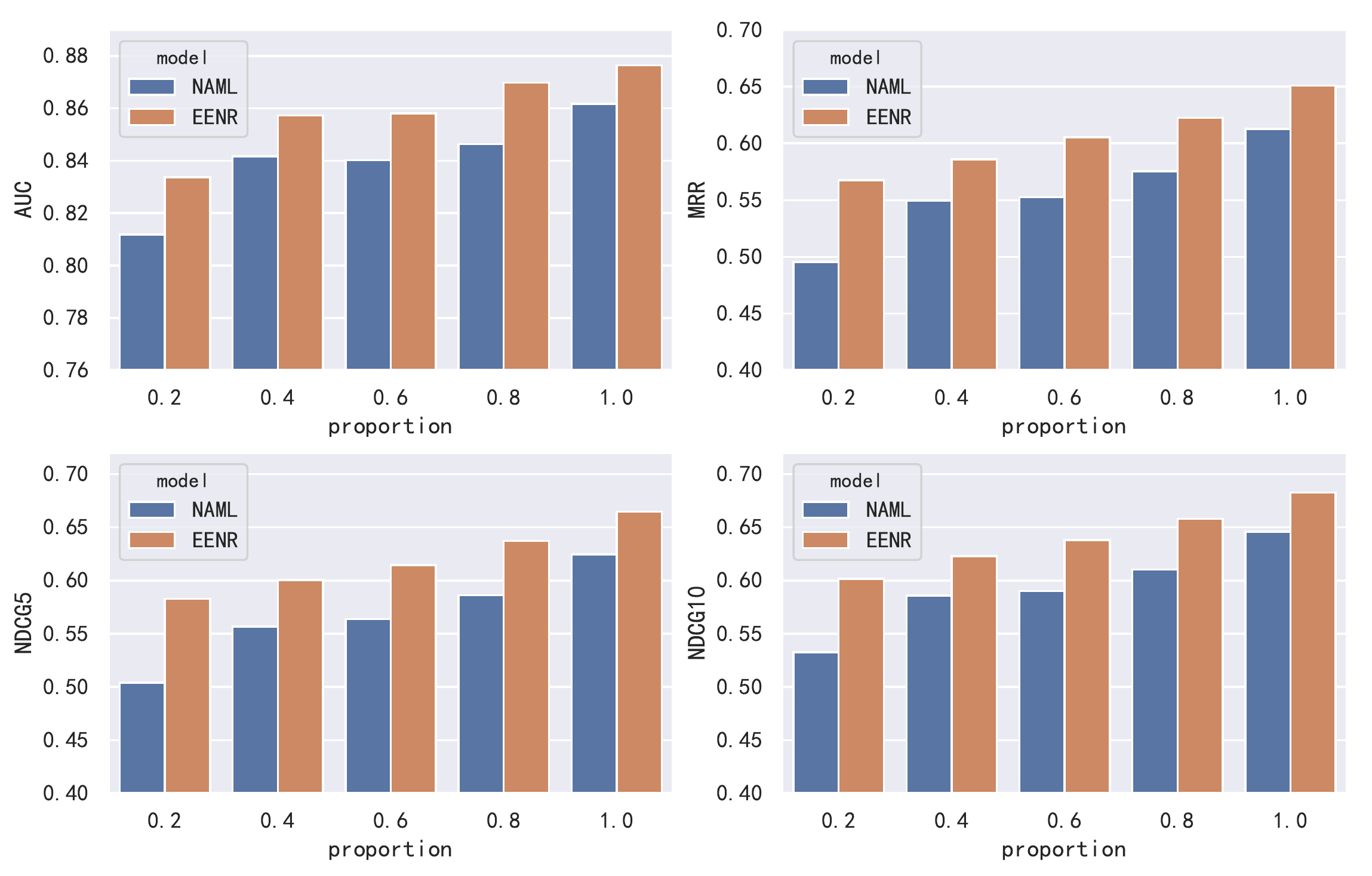}	
	\caption{The performances of EENR compared with NAML in small dataset scenarios.}
	\label{fig4}	
\end{figure}

\paragraph{(3)}  We conduct experiments to verify our speculation that EENR has superiority on small datasets. We randomly select 20\%, 40\%, 60\%, and 80\% of the target news dataset to train and evaluate the model performance. We repeat every model 10 times and report the results in Table 2 (20\%, 60\% of news recommendation dataset). From Table 2, EENR obtains significantly better average AUC (8\%), MRR (7\%), NDCG@5 (8\%), and NDCG@10 (7\%) improvement than NAML when 20\% percent of the total dataset is used. The corresponding figures are 1.5\%, 3.6\%, 3.7\%, 3.4\% when the total dataset is used. More detailed comparisons with NAML at different proportions are plotted in Figure \ref{fig4}, which demonstrates that EENR has better superiority in small dataset scenarios. We attribute the superiority to the event extraction module. The uniqueness of the event extraction component is that it is trained once and applied to any other similar datasets. This separated training strategy leads to fewer hyper-parameters in the subsequent modules of EENR, especially in news body representation. It is more sensible for small datasets that may suffer under-fitting when we use too many parameters to weigh the news body words.

\subsection{Discussion}
This section discusses why the event extraction module could enhance recommendation performance.

Before people click certain news, what they see is a series of news headlines. News titles are the most straightforward and crucial information to people’s attention, and this explains why many early researchers only use titles to encode news and user representation. Recent research considers multi-channel information and utilizes artificial neural networks to model this behavior. However, it may still be insufficient because much research mainly focuses on word-level information when considering news content. We review the human neuron's working process. When facing a series of news headlines, a set of neurons would respond to the optic nerve signals. The response includes computing the meanings of the titles and speculating their contents based on the titles and the knowledge stored in people's brains. People click certain news after a set of neurons completing “calculations”. After Reviewing this process, we get two observations. First, the "contents" before people clicking the news are abstract higher-level (e.g., event-level), not in detail (e.g., word-level). Second, neurons utilize other information stored in brains coming from people’s life experiences.

Most previous research ignores the first observation above, uses content words to encode news bodies. We argue that all words in the news body participation are unsuitable as model input, which bring more noise. As to the second observation, researchers recently began utilizing knowledge graphs to enhance external knowledge in modeling.

We adopt two strategies to narrow these two gaps. First, we use event extraction to abstract higher-level information because EE is naturally designed to answer the 5W1H questions. Second, we use the hot search board dataset to train the event extraction module. External knowledge is stored in the network structure and its parameters. We argue that the EENR framework is closer to real human brain activities than other content-based baselines, thus getting the best performance.

\section{Conclusion}
In this paper, we propose a neural news recommendation framework EENR. EENR contains an event extraction module that extracts abstract higher-level information, such as event arguments, roles, and news types, which provide external knowledge and additional perspectives. The uniqueness of the event extraction component is that it is trained once and can be applied to any other similar datasets. This separated training strategy leads to fewer hyper-parameters in subsequent modules of the framework. And it is more sensible for small datasets. The EENR framework integrates multi-channel information, including news titles, original categories, event types, roles, and arguments to encode news and users. Experimental results on a real-world dataset show that our EENR framework outperforms state-of-the-art benchmarks. Finally, we also conduct experiments to verify its superiority in small dataset scenarios and discuss the rationality to use event-level information to substitute news body content. Future work includes: (1) Improving the accuracy of event extraction. (2) Considering introducing open domain event extraction methods, which help detect unseen event types, to enhance the generalization of the event extraction module.

\bibliographystyle{unsrt}  
\bibliography{references}

\end{document}